\documentclass[pra,aps,floatfix,showpacs,tightenlines,superscriptaddress,amsmath,amssymb,showkeys,10pt,nofootinbib]{revtex4-2}
\usepackage{slashed}
\usepackage{xcolor}
\usepackage{graphicx}

\usepackage{hyperref}

\newcommand{\be}{\begin{equation}}\newcommand{\ee}{\end{equation}}
\newcommand{\bea}{\begin{eqnarray}}\newcommand{\eea}{\end{eqnarray}}
\newcommand{\brr}{\begin{array}}\newcommand{\err}{\end{array}}
\newcommand{\bit}{\begin{itemize}}\newcommand{\eit}{\end{itemize}}
\newcommand{\ben}{\begin{enumerate}}\newcommand{\een}{\end{enumerate}}

\newcommand{\bbm}{\begin{bmatrix}}\newcommand{\ebm}{\end{bmatrix}}
\newcommand{\ba}{\begin{array}}
\newcommand{\ea}{\end{array}}
\newcommand{\G}{\textbf}

\definecolor{darkred}{rgb}{.8,0,0}

\definecolor{darkblue}{rgb}{0,0,0}

\newtheorem{mydef}{Definition}
\newtheorem{Lemma}{Lemma}
\newcommand{\bd}{\begin{mydef}} \newcommand{\ed}{\end{mydef}}
\newcommand{\bthe}{\begin{theorem}} \newcommand{\ethe}{\end{theorem}}
\newcommand{\ble}{\begin{Lemma}} \newcommand{\ele}{\end{Lemma}}

\newcommand{\dr}{\mathrm{d}}

\def\ha{\frac{1}{2}}

\def\intx{\int \!\!\mathrm{d}^3 {\G x}}

\def\lab{\label}\def\lan{\langle}
\def\lf{\left}

\def\non{\nonumber}\def\pa{\partial}\def\ran{\rangle}
\def\rar{\rightarrow}
\def\ri{\right}
\def\al{\alpha}\def\bt{\beta}\def\ga{\gamma}
\def\de{\delta}\def\De{\Delta}

\def\si{\sigma}
\def\om{\omega}

\def\mass{{_{1,2}}}
\def\flav{{e,\mu}}\def\1{{_{1}}}\def\2{{_{2}}}

\newcommand{\ide}{1\hspace{-1mm}{\rm I}}

\def\noHe0{:\;\!\!\;\!\!:H_L(0):\;\!\!\;\!\!:}
\def\noHm0{:\;\!\!\;\!\!:H_\mu(0):\;\!\!\;\!\!:}

\def\lab{\label}
\def\lan{\langle}
\def\lf{\left}

\def\non{\nonumber}
\def\pa{\partial}\def\ran{\rangle}
\def\rar{\rightarrow}
\def\ri{\right}

\def\al{\alpha}\def\bt{\beta}\def\ga{\gamma}
\def\de{\delta}\def\De{\Delta}

\def\si{\sigma}
\def\om{\omega}

\def\mass{{_{1,2}}}
\def\flav{{e,\mu}}\def\1{{_{1}}}\def\2{{_{2}}}

\def\Uk{\bar{U}_k}
\def\Vk{\bar{V}_k}
\begin{document}
\title{Quantum field theory of massive chiral fields}
\author{Massimo~Blasone}
\email{blasone@sa.infn.it}
\affiliation{Dipartimento di Fisica, Universit\`a di Salerno, Via Giovanni Paolo II, 132 84084 Fisciano, Italy \& INFN Sezione di Napoli, Gruppo collegato di Salerno, Italy}
\author{Petr Jizba}
\email{p.jizba@fjfi.cvut.cz}
\affiliation{FNSPE, Czech Technical
University in Prague, B\v{r}ehov\'{a} 7, 115 19 Praha 1, Czech Republic\\}
%
\author{Luca~Smaldone}
\email{lsmaldone@sa.infn.it}
\affiliation{Dipartimento di Fisica, Universit\`a di Salerno, Via Giovanni Paolo II, 132 84084 Fisciano, Italy \& INFN Sezione di Napoli, Gruppo collegato di Salerno, Italy}
%
\vspace{3mm}

\begin{abstract}
We present a quantum-field-theoretic treatment of massive chiral fields in which particles possess well-defined chirality and helicity. This framework reproduces the chiral oscillation formula previously obtained in first-quantized approaches and provides a consistent description of weak-interaction processes. We further derive corresponding chiral-energy uncertainty relations.
\end{abstract}


\maketitle

\section{Introduction}
Chiral oscillations have recently attracted considerable attention \cite{DeLeo:1996gt,fukugita2003physics,Bernardini:2005wh,Bernardini:2005cu,Bernardini:2006bq,Bernardini:2006cy,Bernardini:2006cn,Bernardini:2007ew,Bernardini:2007uf,Bittencourt:2020xen,suekane2021quantum,SalimAdam:2021suq,Bittencourt:2022hwn,Li:2023iys,Morozumi:2025gmw}. 
The physical origin of this phenomenon is deeply connected to the parity-violating structure of weak interactions. Charged-current weak processes couple exclusively to left-chiral fermion fields and right-chiral antifermion fields \cite{PhysRevLett.19.1264,Salam:1968rm,pal2014introductory}. Consequently, particles produced at a weak-interaction vertex are created in states of definite chirality. However, for massive fermions chirality is not conserved under free time evolution, since the mass term couples the left- and right-chiral components of the field. As a result, a fermion produced in a state of definite chirality does not remain in that chirality eigenstate during propagation. This mismatch between the chirality selected at production and the subsequent Hamiltonian evolution gives rise to chiral oscillations.

Beyond their conceptual significance, chiral oscillations may also have phenomenological consequences. In particular, it has been suggested that they could leave observable signatures in the cosmic neutrino background \cite{Ge:2020aen,Bittencourt:2020xen}, as well as in electronic transport phenomena in graphene-based systems, where chiral symmetry may be explicitly broken by external potentials \cite{PhysRevB.102.205404}.

Chiral oscillations have been mainly discussed in the framework of first quantization formalism, and only more recently a quantum field approach has been developed, both by means of a perturbative  and non-perturbative treatments.
In the perturbative treatment, the mass term of a Dirac field has been treated as an interaction between the left and right chiral modes \cite{Blasone:2025hjw}
A quantum field theoretic approach to chiral oscillations has been developed in Ref. \cite{Bittencourt:2024yxi}, where chiral charges are diagonalized using an appropriate Bogoliubov transformation. It is shown that the Hilbert space of chiral states is unitarily inequivalent to the conventional Fock space of energy eigenstates. The chiral oscillation probability is then derived as the expectation value of the chiral charge in a state with definite chirality. The result matches the standard chiral oscillation probability obtained in relativistic quantum mechanics.



 Criticism to the physical reality of chiral oscillations have been raised \cite{Smirnov:2025wax,Akhmedov:2025enm}:  in particular, in Ref.\cite{Smirnov:2025wax} it was argued that neutrino chiral oscillations in vacuum would not occur when the details of the production mechanism are properly taken into account. 

 This is an important issue which however could not be accounted for by means of the formalism developed in Ref.\cite{Bittencourt:2024yxi,Blasone:2026fbu}. The reason is that the quantization scheme there introduced only describes the modes of chiral massive field which have the ``good helicity", i.e. the modes for which helicity and chirality coincide in the massless limit.
However, in a weak decay process mediated by charged current, like the one for pion decay, one has invariably both combinations of helicity and chirality appearing. Thus, it is necessary  to extend the formalism of Ref.\cite{Bittencourt:2024yxi} in order to deal with these situations.

In this paper, we present a complete formalism for chiral massive (Dirac) fields in the context of quantum field theory.
This is done by identifying as a complete set of commuting observables   helicity and  chirality, rather than helicity and energy, as usually done.
Physical states are then identified as eigenstates of the chiral charges (with fixed helicity). Since chiral charges and Hamiltonian do not commute, one has chiral-energy uncertainty relations, which pose constraints on the precision with which one could measure energy for a particle created in a weak-interaction vertex.

Our formalism allows us to calculate at once both chiral oscillations from expectation values of the chiral charges and to derive the amplitude for a weak decay process. In particular, in this paper we consider the decay $\pi^- \rightarrow l^- \, + \, {\bar \nu}_l$ and show that indeed chiral oscillations of the decay products are predicted.

The paper is organized as follows. In Section II we briefly review the chiral oscillations in a first-quantized formalism. Second quantization of the massive chiral fields $\psi_R$ and $\psi_L$ is presented in Section III. This formalism is then applied to the derivation of chiral oscillations and the calculation of the pion decay amplitude. Finally, in Section IV, the chirality-energy uncertainty relations are discussed.

\section{Chiral oscillations in weak decays}
In a charged current weak decay (let us limit to one generation of leptons), the lepton current has the form
\be
J_l^{\mu}(x) \ = \ \bar{\nu}_L(x) \, \ga^\mu \, e_L(x) + h.c. \ , 
\ee
where $\nu(x)$ and $e(x)$ are the neutrino and the electron field, respectively. The chiral fields are defined by
\be \label{leftright}
\psi_{L (R)}(x) \ = \ P_{L (R)} \, \psi(x) \, , 
\ee
where $\psi(x)$ can be either the neutrino or the electron field, and 
\be
P_L \ \equiv \ \ha (1-\ga^5) \, , \qquad P_R \ \equiv \ \ha (1+\ga^5) \, ,
\ee
are the left and right-chiral projectors. Then, only left-chiral fermion fields are involved in weak interactions. The   (effective) interaction Lagrangian can be  written in a Fermi current-current form 
\be
\mathcal{L}_{int} \ = \ \frac{G_F}{\sqrt{2}} \, J^\mu_l \, J_{q,\mu} \, .
\ee
where $J_{q,\mu} $ is the quark part, which will not be explicitly computed in this work. 

Therefore, the amplitude of a process of the type
\be\label{weakdecay}
X \to Y + l+ \overline{\nu}_l  \, , 
\ee
 such process can be computed as
\be \label{tevev}
\lan l_L \, \overline{\nu}_l \, Y|U(t_i,t_f)|X\ran \ \approx \ \ide- i  \frac{G_F}{\sqrt{2}} \, \int^{t_f}_{t_i} \dr^4 x  \, \lan Y| J_{q,\mu}|X\ran \, \lan l_L \, \overline{\nu}_l|J^{\mu}_l|0\ran \, ,  
\ee
where $U(t)$ is the time-evolution operator and we stopped at the first order in  the Dyson expansion.

For simplicity, here we take the neutrino as massless: thus its state is well-defined as excitation of a Weyl field.
What is then not obvious is how to define $|l_L\ran$. 
We will come back to this question in \S \ref{QFTchiral}.

\subsection{Chiral oscillations in QM}
Since chirality is not conserved for massive particles, the chirality of a fermion, such as the electron in the above example, can change between interactions. If we were to measure it in a subsequent process, we might observe an electron flux smaller than expected. This effect is known as \emph{chiral oscillation}.

We start from the Dirac Hamiltonian in first-quantization language
\be \label{diracham}
H \ = \ \al_3 \, p_z+ m \, \beta \, , 
\ee
where $\al$ and $\beta$ are the Dirac matrices
Consider, e.g. an electron, is produced in an initial $L$ state and with some helicity $r$. The state of the electron moving along the $z$-axis ($k^\mu=(\om_k,0,0,k)$), in first quantization is described by the spinor
\be
u_{k,L}^r \ = \ N^r_k \, P_L \, u^r_{ k}\, ,  \qquad r=1,2 \, , 
\ee
where $P_L=(1-\ga^5)/2$ is the projector over the left component, and $u^r_{ k, L}$ is a Dirac spinor with definite helicity $N^r_k$ is introduced to properly normalize the spinor after the projection. 
Note that $P_L$ does not commute with the Dirac Hamiltonian \eqref{diracham}, therefore $u_{k,L}^r$ is not an energy eigenstate. This is a crucial point which be  further discussed in the following.

In order to fix the ideas, suppose an electron with chirality $L$ and helicity $-1$ ($r=2$) is produced in a weak decay. We ask what is the probability for such electron to be observed with the same chirality (helicity is conserved) in another charged weak process. In practice, we have to compute the amplitude
\be \label{ampLL}
\mathcal{A}_{L \to L}(t) \ = \ u^{2\dag}_{k,L} \, U(t) \, u^2_{k,L} \, , \qquad U(t) \equiv e^{-i H t} \, .
\ee
Using Eq.\eqref{diracham}, so that
\be
U(t) \ = \ \ide \, \cos (\om_k t)- \frac{i \, H}{\om_k} \,  \sin (\om_k t)\, , 
\ee
one finds
\be
\mathcal{A}_{L \to L}(t) \ = \ \cos (\om_k t) - \frac{i \, k}{\om_k} \, \sin(\om_k t) \, .
\ee
Therefore
\be \label{pll}
P_{L \to L}(t) \ = \ |\mathcal{A}_{L \to L}(t)|^2 \ = \ 1-\frac{m^2}{\om_k^2} \, \sin^2(\om_k t) \, ,
\ee 
which  is the standard chiral oscillation formula \cite{DeLeo:1996gt,Bernardini:2005wh,Bernardini:2005cu,Bittencourt:2020xen,Bittencourt:2022hwn}.
\section{Quantum field theory of massive chiral fields}
\label{QFTchiral}

In this Section we tackle the problem of quantization of massive chiral fields. This was already considered in Refs.\cite{Bittencourt:2024yxi,Blasone:2026fbu} where it was found that ladder operators with definite chirality and helicity can be defined by means of a Bogoliubov transformation. This allowed to reproduce correctly the  chiral oscillation formula and also the combined chiral and flavor oscillations for neutrinos. 

However, a fundamental limitation of the formalism of Refs.\cite{Bittencourt:2024yxi,Blasone:2026fbu} resides in the impossibility of describing field modes with definite chirality and helicity when they are aligned in the opposite way with respect to the ``natural" one, defined by the massless limit, where chirality and helicity coincide. 
In other words, in that formalism one could describe only the {\em ``good" particles}, e.g. particles with right chirality and positive helicity or those with left chirality and negative helicity (and respective antiparticles). The {\em ``bad" particles}, like a particle with right chirality and negative helicity, do not appear in that quantization scheme. Such modes are however necessary for describing weak interaction processes as the one in Eq.\eqref{weakdecay}.

The reason is that such ``bad" field modes are disappearing in the massless limit. The quantization scheme of Refs.\cite{Bittencourt:2024yxi,Blasone:2026fbu}, based on canonical transformations, which are unitary at finite volume, cannot  deal with such a feature and a generalization  is thus in order, which is what is presented in the following.

Let us now consider the Lagrangian of a massive Dirac field 
\be
{\cal L} \ = \ \overline{\psi}  (i \slashed{\pa} - m )\psi \, .
\ee
We can  decompose $\psi$ in terms of its chiral projections \eqref{leftright}
\be \label{fieldec}
\psi(x) \ = \ \psi_L(x)+\psi_R(x) \, .
\ee
The Lagrangian thus reads
\bea 
\mathcal{L} \  = \
= \sum_{\si=L,R} \overline{\psi}_\si i \slashed{\pa} \psi_\si-m \lf(\overline{\psi}_L \psi_R+\overline{\psi}_R \psi_L\ri)\, , 
\label{Linteract}
\eea

We note that the non-diagonal mass matrix of chiral fields can be diagonalized via a field rotation \cite{cheng1984gauge}:
\begin{eqnarray}
    \Psi &=& \frac{1}{\sqrt{2}}\left(\psi_L \, + \, \psi_R\right)
    \\ 
    {\widetilde\Psi} &=&\frac{1}{\sqrt{2}}\left( - \psi_L \, + \, \psi_R\right)
\end{eqnarray}
and the inverse transformation
\begin{eqnarray} \label{psipsitilde1}
    \psi_R &=& \frac{1}{\sqrt{2}}\left(\Psi \, + \, {\widetilde\Psi}\right)
    \\ \label{psipsitilde2}
     \psi_L &=&\frac{1}{\sqrt{2}}\left( \Psi \, - \, {\widetilde\Psi}\right)
\end{eqnarray}
Thus the above Lagrangian  is diagonalized in terms of two independent Dirac fields, with masses $m$ and $-m$:
\bea 
\mathcal{L} \  = \
\overline{\Psi}  (i \slashed{\pa} - m )\Psi \
+ \, \overline{{\widetilde\Psi}}  (i \slashed{\pa} + m ){\widetilde\Psi}, 
\label{LPsiPsitilde}
\eea

Notice that  \cite{cheng1984gauge}  the two fields are related by ${\widetilde\Psi} = \gamma_5 \Psi$ and that
\be \label{sq2}
\Psi(x) \ = \ \frac{1}{\sqrt{2}} \, \psi(x) \, .
\ee
The factor $\sqrt{2}$ is related to the fact  that we applied a non unitary transformation (projection) on the original Dirac field $\psi$.

We impose canonical equal time anticommutation relations for the $\Psi$ and ${\widetilde\Psi}$ fields\footnote{Note that the anticommutation relation for  ${\widetilde\Psi}$ follows from that for $\Psi$.}:
\be
\Big\{ \Psi(t,\G x), \Psi^\dag(t,\G x') \Big\} \ = \ \de^3(\G x-\G x')\ = \ \Big\{ {\widetilde\Psi}(t,\G x), {\widetilde\Psi}^\dag(t,\G x') \Big\}   \, .
\ee

The expansions for these fields   are:
\begin{eqnarray}
\Psi(x) &=&  \sum_{r=1,2} \int\frac{d^3{\bf k}}{(2\pi)^{3/2}}\,  \left[ u_{{\bf k}}^{r}\, e^{-i \om_k t} \, \alpha_{{\bf k}}^{r} \, +\, v_{-{\bf k}}^{r}\, e^{i \om_k t} \, \bt_{-{\bf k}}^{r\dagger} \right]  e^{i{\bf k}\cdot {\bf x}}  \, .
\label{psi}
\\
 {\widetilde \Psi}(x) &=&   \sum_{r=1,2} \int \frac{d^3{\bf k}}{(2\pi)^{3/2}}\,  \left[ {\widetilde u}_{{\bf k}}^{r}\, e^{-i \om_k t} \, {\widetilde \alpha}_{{\bf k}}^{r} \, +\, {\widetilde v}_{-{\bf k}}^{r}\, e^{i \om_k t} \, {\widetilde \bt}_{-{\bf k}}^{r\dagger} \right]  e^{i{\bf k}\cdot {\bf x}}  \, ,
 \label{psitilde}
\end{eqnarray}
with $\omega_k=\sqrt{k^2+m^2}$
and
\bea \label{orth}
u^{r\dag}_{{\bf k},i} u^{s}_{{\bf k},i} =
v^{r\dag}_{{\bf k},i} v^{s}_{{\bf k},i} =  \de_{rs}
\, , \quad u^{r\dag}_{{\bf k},i} v^{s}_{-{\bf k},i} = 0
\,, \quad \sum_{r}(u^{r\l*}_{{\bf k},i} u^{r\bt}_{{\bf k},i} +
v^{r\al*}_{-{\bf k},i} v^{r\bt}_{-{\bf k},i}) =  \de_{\al\bt}\;,
\eea
and similar ones for the tilde spinors
 ${\widetilde u},{\widetilde v}$,  which are obtained from $u,v$ by $m\rightarrow - m$.
 
The creation and annihilation operators satisfy the usual anticommuation relations
\begin{equation}
    \Big\{ \alpha_{\bf k}^r, \alpha_{\bf p}^{s\dagger}\Big\} = 
    \Big\{ \beta_{\bf k}^r, \beta_{\bf p}^{s\dagger}\Big\} = \delta^3({\bf k}-{\bf p}) \delta_{rs} = 
    \Big\{ {\widetilde \alpha}_{\bf k}^r, {\widetilde \alpha}_{\bf p}^{s\dagger}\Big\} = 
    \Big\{ {\widetilde \beta}_{\bf k}^r, {\widetilde \beta}_{\bf p}^{s\dagger}\Big\}\, ,
\end{equation}
and other anticommutators vanishing.

We now proceed in determining the chiral ladder operators with definite chiralities and helicities (but no definite energy) from the above relations.

The above transformations \eqref{psipsitilde1},\eqref{psipsitilde2} 
can be regarded as mixing transformations for two fields with different masses $m_1=-m$ and $m_2=m$ and maximal mixing $\theta=\pi/4$. The chiral fields play the role of the flavor fields. The quantum field theory of (Dirac) fermion mixing in the case of two flavors \cite{Blasone:1995zc} can be thus straightforwardly applied (see Appendix) and we can write the above transformations as
\bea\label{mix1}
\psi_R(x) & = & \bar{G}^{-1}(t) \, \widetilde{\Psi}(x) \bar{G}(t) \, ,
\\ [2mm] \label{mix2}
\psi_L(x) & = & \bar{G}^{-1}(t) \, \Psi(x) \bar{G}(t) \, , 
\eea
where the generator of the field mixing transformation is 
\be
\bar{G}(t) \ = \ \exp\lf[\frac{\pi}{4} \intx \lf(\widetilde{\Psi}^\dag(x) \Psi(x) -\Psi^\dag(x) \widetilde{\Psi}(x)\ri)\ri] \, . 
\ee

The transformation Eqs.\eqref{mix1},\eqref{mix2} acts on creation and annihilation operators as a combination of a rotation and of a Bogoliubov transformation \cite{Blasone:1995zc}, giving
\begin{eqnarray}\label{chiraltransf1a}
    \alpha_{{\bf k},R}^{1}(t) &=&
    \frac{1}{\sqrt{2}}  \widetilde{\alpha}_{{\bf k}}^{1} \,+\,\frac{1}{\sqrt{2}}\left(\Uk\, \alpha_{{\bf k}}^{1}\,+\, \Vk\, \bt_{-{\bf k}}^{1\dagger}e^{2 i \om_k t}\right)
    \\ \label{chiraltransf1b}
    \bt_{-{\bf k},R}^{1\dagger}(t) &=&
     \frac{1}{\sqrt{2}} \widetilde{\bt}_{-{\bf k}}^{1\dagger} \,+\,\frac{1}{\sqrt{2}}\left(\Uk\,  \bt_{-{\bf k}}^{1\dagger}\,-\, \Vk\,
     \alpha_{{\bf k}}^{1}e^{-2 i \om_k t}
     \right)
     \\ \label{chiraltransf1c}
        \alpha_{{\bf k},L}^{1}(t) &=&
    \frac{1}{\sqrt{2}} \alpha_{{\bf k}}^{1}\,-\,\frac{1}{\sqrt{2}}\left(\Uk\, \widetilde{\alpha}_{{\bf k}}^{1} \,-\, \Vk\, \widetilde{\beta}_{-{\bf k}}^{1\dagger}e^{2 i \om_k t}\right)
    \\ \label{chiraltransf1d}
    \bt_{-{\bf k},L}^{1\dagger}(t) &=&
     \frac{1}{\sqrt{2}}  \bt_{-{\bf k}}^{1\dagger} \,-\,\frac{1}{\sqrt{2}}\left(\Uk\, \widetilde{\beta}_{-{\bf k}}^{1\dagger}\,+\, \Vk\,
     \widetilde{\alpha}_{{\bf k}}^{1}e^{-2 i \om_k t}
     \right)
\end{eqnarray}
and
\begin{eqnarray}
\label{chiraltransf2a}
   \alpha_{{\bf k},R}^{2}(t) &=&
    \frac{1}{\sqrt{2}}  \widetilde{\alpha}_{{\bf k}}^{2}\,+\,\frac{1}{\sqrt{2}}\left(\Uk\,\alpha_{{\bf k}}^{2}\,-\, \Vk\, \bt_{-{\bf k}}^{2\dagger}e^{2 i \om_k t}\right)
    \\ \label{chiraltransf2b}
   \bt_{-{\bf k},R}^{2\dagger} (t) &=&
     \frac{1}{\sqrt{2}} \widetilde{\bt}_{-{\bf k}}^{2\dagger} \,+\,\frac{1}{\sqrt{2}}\left(\Uk\,  \bt_{-{\bf k}}^{2\dagger}\,+\, \Vk\,
      \alpha_{{\bf k}}^{2}e^{-2 i \om_k t}
     \right)
     \\ \label{chiraltransf2c}
     \alpha_{{\bf k},L}^{2}(t) &=&
    \frac{1}{\sqrt{2}}   \alpha_{{\bf k}}^{2}\,-\,\frac{1}{\sqrt{2}}\left(\Uk\, \widetilde{\alpha}_{{\bf k}}^{2} \,+\, \Vk\, \widetilde{\beta}_{-{\bf k}}^{2\dagger}e^{2 i \om_k t}\right)
   \\ \label{chiraltransf2d}
   \bt_{-{\bf k},L}^{2\dagger}(t)&=&
     \frac{1}{\sqrt{2}}  \bt_{-{\bf k}}^{2\dagger}\,-\,\frac{1}{\sqrt{2}}\left(\Uk\, \widetilde{\beta}_{-{\bf k}}^{2\dagger} \,-\, \Vk\,
    \widetilde{\alpha}_{{\bf k}}^{2}e^{-2 i \om_k t}
     \right)
\end{eqnarray}
where 
$\Uk$ and $\Vk$ are the (chiral) Bogoliubov coefficients, which are defined as
\begin{eqnarray}\label{uvtilde1}
    {\widetilde u}_{{\bf k}}^{r\dagger}u_{{\bf k}}^{r} =  {\widetilde v}_{-{\bf k}}^{r\dagger}v_{-{\bf k}}^{r}=\frac{k}{\omega_k} \equiv \Uk
    \\ \label{uvtilde2}
   {\widetilde u}_{{\bf k}}^{1\dagger}v_{-{\bf k}}^{1}= -{\widetilde v}_{-{\bf k}}^{1\dagger}u_{{\bf k}}^{1}=  {\widetilde v}_{-{\bf k}}^{2\dagger}u_{{\bf k}}^{2}= - {\widetilde u}_{{\bf k}}^{2\dagger}v_{-{\bf k}}^{2}  = \frac{m}{\omega_k} \equiv \Vk
\end{eqnarray}
with $\Uk^2+\Vk^2=1 $. 

The left and right fields can be thus expanded as 
\begin{eqnarray} \label{psiR}
\psi_R &=&   \sum_{r=1,2}\int\frac{d^3{\bf k}}{(2\pi)^{3/2}} \Big[ 
 {\widetilde u}_{{\bf k}}^{r} \, e^{-i \om_k t}\,\alpha_{{\bf k},R}^{r}(t) 
+ 
{\widetilde v}_{-{\bf k}}^{r} \, e^{i \om_k t}\,\beta_{-{\bf k},R}^{r\dagger}(t) 
\Big]   \, ,
\\ \label{psiL}
\psi_L &=& \sum_{r=1,2} \int \frac{d^3{\bf k}}{(2\pi)^{3/2}}\Big[ 
 u_{{\bf k}}^{r} \, e^{-i \om_k t} \,\alpha_{{\bf k},L}^{r}(t) 
+ 
 v_{-{\bf k}}^{r} \, e^{i \om_k t} \,\beta_{-{\bf k},L}^{r\dagger}(t) \, .
\Big]   \, .
\end{eqnarray}
By inserting the explicit form of the chiral ladder operators \eqref{chiraltransf1a}-
\eqref{chiraltransf2d}, one can check the validity of Eqs.\eqref{mix1},\eqref{mix2} (see Appendix).

The vacuum  annihilated by $\al^r_{\G k,\si}(t)$ and $\bt^r_{\G k,\si}(t)$ is defined as $|0(t)\ran_{L,R}\equiv \bar{G}^{-1}(t) |0\ran$. Its explicit form at $t=0$ can be obtained by the flavor vacuum (see Eq.\eqref{vacuumflav}): 
\begin{eqnarray}
&& \label{vacuumchir}
|0\ran_{L,R}= \prod_{{\bf k},r} \frac{1}{2}\Big[
(2- \,\Vk^2)
+\,\epsilon^r \,\Vk
\, ({\widetilde \al}^{r\dagger}_{{\bf k}}\bt^{r\dagger}_{-{\bf k}}+
\al^{r\dagger}_{{\bf k}} {\widetilde \bt}^{r\dagger}_{-{\bf k}})
-\,\epsilon^r  \,\Uk\Vk \,(
{\widetilde \al}^{r\dagger}_{{\bf k}}{\widetilde \bt}^{r\dagger}_{-{\bf k}} -
\al^{r\dagger}_{{\bf k}}\bt^{r\dagger}_{-{\bf k}} )
+  \, \Vk^2
\, {\widetilde \al}^{r\dagger}_{{\bf k}}\bt^{r\dagger}_{-{\bf k}}
\al^{r\dagger}_{{\bf k}}{\widetilde \bt}^{r\dagger}_{-{\bf
k}}\Big]|0\ran  \, .
\end{eqnarray}
This state is a condensate of particle-antiparticle pairs. In particular, it is an entangled state of modes of the  fields $\Psi$ and ${\widetilde \Psi}$.

The condensation density is given by:
\begin{equation}
{}_{L,R} \lan 0|  \al^{r\dagger}_{{\bf k}}\al^{r}_{{\bf k}}  |0\ran_{L,R} \, = \, {}_{L,R} \lan 0|  {\widetilde \al}^{r\dagger}_{{\bf k}}{\widetilde \al}^{r}_{{\bf k}}  |0\ran_{L,R} \, = \,\frac{1}{2} \Vk^2\, =\, \frac{m^2}{2\,\omega_k^2},
\end{equation}
with the same result for antiparticles. Note that this vanishes in the relativistic limit $k\gg m$.

As already stressed, in Ref.\cite{Bittencourt:2024yxi} operators with definite chirality and helicity have been constructed out of the $\psi$ field only, thus resulting in only ``half" of the operators given above. It is interesting to note, for example,  that the term in round brackets  in Eq.\eqref{chiraltransf1a} is formally the same as the operator $\alpha_R$ of Ref.\cite{Bittencourt:2024yxi}: thus it is an interesting question to ask  about the relation between the two formalisms.

\subsection{Chiral charges, chiral states and the oscillation formula}
Chiral charges are defined by
\begin{equation}
    Q_{L}(t) =  \int d^3{\bf x}\, \psi_{L}^\dagger (x)\psi_{L}(x), \qquad  Q_{R} (t)= \int d^3{\bf x}\, \psi_{R}^\dagger(x) \psi_{R}(x).
\end{equation}
Note also that the total (conserved) charge is equal to the sum of the (time dependent) chiral charges
\begin{equation}
    Q  \, = \,Q_{L}(t) \, +\, Q_{R}(t).
\end{equation}
Let us observe the important fact that chiral charges are time dependent, and then do not commute with the Hamiltonian. In other words, a state with a definite chirality has no-definite energy, unless $m=0$.
The chiral charges are diagonalized in terms of the above chiral ladder operators:
\begin{eqnarray}\label{Qchiral}
    Q_\si(t)&=& \sum_{r=1,2}\int d^3{\bf k} \,\left(\alpha_{{\bf k},\si}^{r \dagger}(t)\alpha^r_{{\bf k},\si}(t)
    - \,\beta_{-{\bf k},\si}^{r \dagger}(t)\beta^r_{-{\bf k},\si}(t) \right), \quad \si = L,R.
\end{eqnarray}

We then define the state of a fermion with a definite chirality and helicity as
\be
|\psi^r_{\G k,\si}(t)\ran \ = \ \sqrt{2 \om_k (2 \pi)^3} \,  \al^{r \dag}_{\G k,\si}(t)|0(t)\ran_{L,R} \, , 
\ee
so it has the covariant normalization 
\be
\lan \psi^r_{\G k,\si}(t)|\psi^s_{\G p,\si}(t)\ran \ =  \ 2 \om_k \, (2 \pi)^3 \, \de^3(\G k-\G p) \, .
\ee

We now want to write down the time evolved chiral ladder operators in terms of the ones at $t=0$. Because the Hamiltonian is quadratic their evolution will be described by a linear combination of the form
\bea
\al^r_{\G k,\si}(t) & = & \sum_{\rho=L,R} \, \lf(f^{\G k}_{\si \rho}(t) \, \al^r_{\G k,\rho} + g^{\G k,r}_{\si \rho}(t)\,  \bt^{r\dag}_{-\G k,\rho}\ri) \, , \\[2mm]
\bt^{r\dag}_{-\G k,\si}(t) & = & \sum_{\rho=L,R} \, \lf(-g^{\G k,r *}_{\si \rho}(t)\,  \al^r_{\G k,\rho} + f^{\G k *}_{\si \rho}(t) \, \bt^{r\dag}_{-\G k,\rho}\ri) \, ,
\eea
where the operators without argument are evaluated at $t=0$ and we have defined
\be
f^{\G k}_{\si \rho}(t-t_i) \ = \ \lf\{\al^r_{\G k,\si}(t),\al^{r\dag}_{\G k,\rho}(t_i)\ri\} \,, \qquad g^{\G k,r}_{\si \rho}(t,t_i) \ = \ \lf\{\al^r_{\G k,\si}(t),\bt^{r}_{-\G k,\rho}(t_i)\ri\} \, .
\ee
and $g^{\G k}_{\rho,\si}(t) \equiv g^{\G k}_{\rho,\si}(t,0)$.
Explicitly
\bea
f^{\G k}_{\si \si}(t-t_i) & = & \ha \lf(1+ \Uk^2 + \Vk^2 \, e^{2 i \om_k (t-t_i)}\ri) \, ,  \qquad \si=L,R \, , \\[2mm]
f^{\G k}_{\si \rho}(t) & = & 0 \, ,  \qquad \si \neq \rho \, , \\[2mm]
g^{\G k,r}_{L L}(t,t_i) & = & -g^{\G k,r}_{R R}(t,t_i) \ = \  \ha \, \varepsilon^r \, \Uk \, \Vk \,\lf( e^{2 i \om_k t}- e^{2 i \om_k t_i}\ri) \, , \\[2mm] \label{Gsirho}
g^{\G k,r}_{\si \rho}(t,t_i) & = &   \ha   \varepsilon^r \Vk \,\lf(e^{2 i \om_k t_i}-e^{2 i \om_k t}\ri) \, , \qquad \si \neq \rho \, ,
\eea 
with $\varepsilon^r \equiv (-1)^r$.
Notice that such relations are independent on helicity, as it should be, because helicity is conserved in the time-evolution. Using such relations, e.g., one can rewrite the expansion \eqref{psiL} as
\be \label{psiti}
\psi_L(x) \ = \ \sum_r \, \sum_{\rho=L,R}  \int \frac{d^3{\bf k}}{(2\pi)^{3/2}} \, \lf(U^{\G k,r}_{L \rho}(t,t_i) \al^r_{\G k,\rho}(t_i)+V^{-\G k,r}_{L \rho}(t,t_i) \bt^{r\dag}_{-\G k,\rho}(t_i) \ri) \, e^{i \G k \cdot \G x} \, , \qquad \si \ = \ L,R \, ,
\ee
where
\bea
U^{\G k, r}_{L \rho}(t,t_i) & = &  u^r_{\G k} e^{-i \om_k t} f^{\G k}_{L \rho}(t-t_i)-v^r_{-\G k} e^{i \om_k t} g^{\G k,r*}_{L \rho}(t,t_i) \, . \\[2mm]
V^{-\G k, r}_{L \rho}(t,t_i) & = & v^r_{-\G k} e^{i \om_k t} f^{\G k *}_{L \rho}(t-t_i)+u^r_{\G k} e^{-i \om_k t} g^{\G k, r}_{L \rho}(t,t_i) \, .
\eea
The expansion \eqref{psiR} can be rewritten in a similar way, but the tilde-spinors will appear.

The oscillation probability will be thus given by ($\si \neq \rho $)
\bea
&& P_{\si\rightarrow \rho}(t) \ = \ \lan :Q_\rho (t): \ran_\si  \ = \ |g^{\G k,r}_{\si \rho}(t)|^2\, , 
\\[2mm]
&& P_{\si\rightarrow \si}(t)  \ = \  1 \ - \ P_{\si\rightarrow \rho}(t) \ = \ |f^\G k_{\si \si}(t)|^2+ |g^{\G k,r}_{\si \si}(t)|^2 \,  ,
\eea
where we have the defined the average as
$
 \lan \ldots \ran_\si \equiv  \frac{\lan \psi^r_{\G k,\si}| \ldots|\psi^r_{\G k,\si}\ran}{\lan \psi^r_{\G k,\si}|\psi^r_{\G k,\si}\ran} 
$,
and the normal ordering is meant with respect to the operators at $t=0$, because we are working in that Hilbert space.
The formula we obtain is the same as in the QM case (see Eq.\eqref{pll}). Notice that the result is independent on the helicity, because the dependence on $g$ only consists in a sign.
 For more formal details see Ref. \cite{Bittencourt:2024yxi}. Note this is independent on the helicity of the initial state, which is conserved along the propagation.

\subsection{Pion decay in terms of chiral states}
Let us consider the pion decay $\pi^- \to \bar{\nu}_e + e^-$. The neutrino field can be taken as massless. 
Then, the antineutrino should have positive helicity and therefore be described by the spinor $v_{-\G p,L}$,
where we expanded the left-handed (Weyl) neutrino field as 
\begin{eqnarray}  
\nu_L(x) &=& \int \frac{d^3{\bf p}}{(2\pi)^{3/2}} \Big[u_{{\bf p},L} \, a^2_{{\bf p}}\,e^{ -i p t}\, +\, v_{{-\bf p},L}\, b_{{-\bf p}}^{1 \dagger}\,e^{ i p t}  \Big] e^{i {\bf p}\cdot{\bf x}} \, , 
\end{eqnarray}
where $u_{\G p,L} \equiv P_L u^2_{\G p}|_{m=0}$ and  $v_{-\G p,L} \equiv P_L v^1_{-\G p}|_{m=0}$-
Therefore, we can compute that the $W^+$ part of the current, which is the one relevant in this computation, acts on the vacuum at $t=0$ as\footnote{In  this section, our notation for the chiral vacuum includes the tensor product with the  neutrino vacuum state.}
\bea \non
J^{+,\mu}_l|0(t_i)\ran_{L,R} & =&
\frac{1}{(2 \pi)^3}\ \sum_{r=1,2} \,\sum_{\si=L,R} \int \dr^3 \G k  \int \!\dr^3 \G p \, e^{i (\G p-\G k) \cdot \G x} \,e^{i p t} \,\overline{U}^{\G k,r}_{L \si}(t,t_i) \ga^\mu v_{-\G p}\, \al^{r \dag}_{\G k,\si}(t_i) \, b^{1\dag}_{-\G p}  |0(t_i)\ran_{L,R} \, \\ [2mm] \non
& = & 
\frac{1}{(2 \pi)^3} \
 \sum_{r=1,2} \,\sum_{\si=L,R} \int \dr^3\! \G k  \int \dr^3 \G p \, e^{i (\G p-\G k) \cdot \G x} \, \lf( e^{i (\om_k+p) t}  f^{\G k *}_{L \si}(t-t_i) \, \overline{u}^r_{\G k}- e^{i (p-\om_k) t}  g^{\G k,r}_{L \si}(t,t_i) \, \overline{v}^r_{-\G k}\ri) \times
 \\ \label{currac}
 &&
 \qquad \qquad \qquad \qquad\qquad \qquad \qquad \qquad
\times \, \ga^\mu \, v_{-\G p,L} \, \al^{r \dag}_{\G k,\si}(t_i) \, b^{1\dag}_{-\G p} |0(t_i)\ran_{L,R} \, .
\eea
Let us now compute the matrix element of the current
\bea \non
 {\cal J}^\mu_{l}(x;t_i) & = & {}_{L,R}\lan 0(t_i)| b^1_{\G p} \al^1_{\G k,L} \, J^{+\mu}_l(x)|0(t_i)\ran_{L,R} \\[2mm] 
& = &  \frac{e^{-i (\G p+\G k) \cdot \G x}}{(2 \pi)^3}   \,  (e^{i (\om_k+p) t} f^{\G k *}_{L L}(t-t_i)  \,  \overline{u}^1_{\G k}-e^{i (p-\om_k) t} g^{\G k,1 }_{L L}(t_f,t_i)  \,  \overline{v}^1_{-\G k}) \, \ga^\mu \, v_{\G p,L}  \, .
\eea
We get
\be
 {\cal J}^\mu_{l}(x;t_i)  \ = \  \frac{e^{-i (\G p+\G k) \cdot \G x}}{2 (2 \pi)^3} \, a^- \, \lf[e^{i  (p+\om_k )t}   \lf(1+\Uk+2\Uk^2\ri) +e^{i  (p-\om_k )t} \, e^{2 i \om_k t_i}  \, \lf(\Vk^2-\Uk^2-\Uk \ri)   \ri] \, ,
\ee
where we have defined
\be \label{apm}
a^\pm \ = \  \frac{k \pm \om_k \pm m}{2 \sqrt{\om_k(m+\om_k)}}
\ee

Here we are describing situations where both production and detection are performed via charged current weak interactions, thus involving pure chiral fermions. As we have seen, the corresponding Fock states, which are eigenstates of chiral charges are intrinsically time-dependent and they are not asymptotically stable states. However, they permit to describe what happens close to the interaction vertex, at $t=0$. Then, we proceed with a finite-time description of the pion-decay, in terms of the time-evolution operator instead of an $S$-matrix description.
We thus define 
\be
\lan \overline{\nu}^1_{-\G p,L} \, l^{-,1}_{\G k} | U(t_f,t_i)-\ide|\pi^-_{\G p_{\pi}} \ran \ = \ i (2 \pi)^3 \de^3(\G p_\pi-\G k-\G p) \, \mathcal{M}(t_i,t_f) \, ,
\ee
where the states are defined at $t_i$.
As usual, we write the pion current matrix element as $J^\mu_\pi= f_\pi p_{\pi}^\mu$. Therefore, employing Eq.\eqref{tevev} we can write, in the pion rest frame
\be
\mathcal{M}(-T,T) \ = \ \frac{G_F \, f_\pi \, m_\pi}{\sqrt{2}} \, \int^{T}_{-T} \dr^4 x  {\cal J}^\mu_{l}(x;-T) \, .
\ee
Explicitly
\be \label{MTT}
\mathcal{M}(-T,T) \ = \  \mathcal{M}^-_{fi}  \, \lf[\frac{\sin[(m_\pi-k-\om_k )T]}{m_\pi-k-\om_k }   \lf(1+\Uk+2\Uk^2\ri) +\frac{\sin[(m_\pi-k+\om_k )T]}{m_\pi-k-\om_k }  \, e^{-2 i \om_k T}  \, \lf(\Vk^2-\Uk^2-\Uk \ri)   \ri] \, .
\ee
where we indicated 
\be
\mathcal{M}^\pm_{fi} \ = \ \sqrt{2 \om_k} \sqrt{2 k} \frac{G_F}{\sqrt{2}} \, f_\pi \, m_\pi \, a^\pm \, . 
\ee
Notice that $\mathcal{M}^-_{fi}$ is the usual invariant matrix element in the energy eigenstates basis. For $T \to \infty$, the second piece on the r.h.s. of Eq.\eqref{MTT} can be neglected because is a strongly oscillating term, and we find
\be
\mathcal{M}(-T,T) \ \approx \ 2 \pi \,  \mathcal{M}^{-}_{fi}  \, \de(m_\pi-k-\om_k ) \, \frac{1}{2} \,  \lf(1+\Uk+2\Uk^2\ri) \, .
\ee
which ensures energy conservation, as required by time-energy uncertainty principle (see \S  \ref{chiralenergy}).

At finite-time, where we do not have the strict energy conservation, is more physical to introduce the amplitude of the process per unit time
\be
\mathcal{A}(-T,T) \ \equiv  \ \frac{\mathcal{M}(-T,T)}{2 T} \, . 
\ee
If we consider $T \ll 1$ (however it should be taken larger than the $W$-boson decau time scale \cite{Blasone:2006jx}), we get
\be
\mathcal{A}(0,0) \ \approx \  \mathcal{M}^-_{fi}  \, .
\ee

Let us now consider the process where a \emph{right-handed} charged lepton is observed. It is interesting to notice that, in such case
\be
 {\cal J}^\mu_{l}(x;t_i) \ = \ {}_{L,R}\lan 0(t_i)| b^1_{\G p} \al^1_{\G k,R} \, J^{+\mu}_l(x)|0(t_i)\ran_{L,R} \ = \  - \, 2 i\, \sqrt{p \, \om_k}\frac{e^{-i (\G p+\G k) \cdot \G x}}{(2 \pi)^3} \, e^{i p t}\, e^{i \om_k t_i}\mathcal{A}_{L \to R}(t-t_i)  \,  a_+ \, ,    \, .
\ee
where $a^+$ has been defined in Eq.\eqref{apm}. $\mathcal{A}_{L \to R}(t-t_i) $ is the chiral oscillations amplitude, so it is clear already at the level of the matrix element of the lepton current that the above process is a consequence of chiral oscillations: right-chiral leptons are not produced in the interaction vertex.

Proceedings as above, we get
\be \label{MTT}
\mathcal{M}(-T,T) \ = \ m \, \sqrt{\Uk} \, \mathcal{M}^+_{fi}  \, \lf[\frac{\sin[(m_\pi-k+\om_k )T]}{m_\pi-k-\om_k }  \, e^{2 i \om_k T}  \,   -\frac{\sin[(m_\pi-k-\om_k )T]}{m_\pi-k-\om_k }   \ri] \, .
\ee

In such case, the behavior close to the vertex gives
\be
\mathcal{A}(0,0) \ \approx \ 0 \, .
\ee
where we used $\approx$ to indicate that $T \ll 1$, but larger than interaction time-scale, as above. 
This is expected, because, as remarked above, right-handed leptons are not produced in our interaction, but they appear because of chiral oscillations.
\section{Chirality-energy uncertainty relations}
\label{chiralenergy}
In the previous section we have built a finite-time QFT formalism, which is necessary to describe chiral states involeved in weak interactions. Because they are eigenstates of the chiral charges, which are no separately conserved, they necessarily have to be described by a finite-time formalism and they do not have definite energy. The situation is similar to the one of unstable particles, where it is known that energy uncertainty and particle lifetime are intimately related by time-energy uncertainty relation. In that case it is well known that a finite-time description is more appropriate, in order to encode short-time effects as the quantum Zeno effect \cite{PhysRevLett.71.2687,facchi1999regola,Giacosa:2011xa,Giacosa:2010br,Giacosa:2018dzm,Giacosa:2021hgl}. Here we study the time-energy uncertainty relation for chiral states, which formalizes such arguments. A similar analysis has been also performed in the case of flavor neutrino states \cite{Bilenky:2005hv,Bilenky2007,Bilenky2008,Blasone:2018ktu,Blasone2020}.

Mandelstam--Tamm version of time-energy uncertainty relation (TEUR) is formulated as~\cite{ManTam}
\be \label{teunc}
\Delta E \, \Delta t \, \geq \frac{1}{2} \, ,
\ee
where
\be
\Delta E \equiv \si_H \, \qquad \Delta t \equiv \si_O/\lf|\frac{\dr \lan O(t) \ran}{\dr t}\ri| \, .
\label{teunc1}
\ee
Here $O(t)$ represents the ``clock observable'' whose dynamics quantifies temporal changes in a system
and $\Delta t$ is the characteristic time interval over which the mean value of $O$ changes by a standard deviation.

Let us apply the above considerations to the first-quantized approach to chiral-oscillations. The best candidate for the clock-observable is the chiral projector over the chiral state $\psi_L$
\be
O(t) \ = \ \Pi_L(t) \ = \ U^\dag(t) \,  u^{2}_{k,L}  \, u^{2\dag}_{k,L} \, U(t) \, .
\ee
Notice that the action of $U(t)$ does not give a simple phase factor, because $H$ does not commute with $\ga_5$. One can easily see that
\be
\Pi^2_L(t) \ = \ \Pi_L(t) \, , 
\ee
and that
\be
P_{L \to L}(t) \ = \ u^{\dag}_{k,L} \, \Pi_L(t) \, u_{k,L}(t) \, .
\ee
Therefore
\bea \label{varq}
\si^2_\Pi & = & P_{L \to L}(t)  (1-P_{L \to L}(t) )  
= \frac{m^2 \sin^2( \omega_\G k t ) \left(\omega^2_\G k-m^2 \sin^2( \omega_\G k t)\right)}{\omega_\G k^4}
\eea
The explicit form of $\Pi_L(t)$ is
\be
\Pi_L(t) \ = \ \left(
\begin{array}{cccc}
 0 & 0 & 0 & 0 \\
 0 & P_{L \to L}(t) & 0 & \frac{m \, k}{\om_k^2} \sin^2 (\om_k t) - i \frac{m}{2 \om_k} \sin (2\om_k t)  \\
 0 & 0 & 0 & 0 \\
 0 & \frac{k}{\om_k}\sin^2 (\om_k t) +\frac{i}{2} \sin(2 \om_k t)  & 0 & 1- P_{L \to L}(t) \\
\end{array}
\right) 
\ee

What about QFT? In that case, chiral charge is a natural candidate for a ``clock observable'', so we can take $O(t)=Q_\si(t)$. Moreover, we will use the average $\lan \ldots \ran_\si$ defined in the previous sections, on the state at the reference time $t=0$.

In any case we get the {\em chiral--energy} uncertainty relation
\be \label{neutun}
\sigma_H \, \sigma_Q \ \geq \ \frac{1}{2}\lf|\frac{\dr P_{\si\rightarrow \si}(t)}{\dr t}\ri| \, , 
\ee
where $\sigma^2_Q \, = \, \lan Q^2_{\si}(t)\ran_\si \ - \ \lan Q_{\si}(t)\ran_\si^2 $ gives the same result as in Eq.\eqref{varq}.
Therefore
\be \label{neutunqm}
\Delta E  \,\sqrt{P_{\si\rightarrow \si}(t)\lf(1-P_{\si\rightarrow \si}(t)\ri)} \ \geq \ \ha \, \lf|\frac{\dr P_{\si\rightarrow \si}(t)}{\dr t}\ri|  \, .
\ee
Then, we have to explicitly compute $\De t$. Because
\be
\frac{\dr P_{\si\rightarrow \si}(t)}{\dr t} \ = \ \frac{m^2}{\om_\G k} \sin(2 \om_\G k t)
\ee
and using Eq.\eqref{varq}, we get
\be
\De t \ = \ \frac{\sqrt{\sin ^2( \omega_{\G k} t ) \left(\omega_{\G k} ^2-m^2 \sin ^2(\omega_{\G k} t )\right)}}{m \omega_{\G k}  | \sin (2  \omega_{\G k} t )| } \, ,
\ee
Then the uncertainty relation looks
\be
\De E \, \frac{\sqrt{\sin ^2( \omega_{\G k} t ) \left(\omega_{\G k} ^2-m^2 \sin ^2(\omega_{\G k} t )\right)}}{m \omega_{\G k}  | \sin (2  \omega_{\G k} t )| }  \geq 1/2 \, .
\ee
If we integrate both members over the first oscillation period, we can take care of the absolute value
\be
\De E \, \int^{\frac{\pi}{2 \om_{\G k}}}_0  \, \dr t \, \frac{\sqrt{ \left(1-\frac{m^2}{\omega_{\G k} ^2} \sin ^2(\omega_{\G k} t )\right)}}{ \frac{m}{\omega_{\G k}}    \cos ( \omega_{\G k} t ) }  \geq \frac{\pi}{2}
\ee

In the relativistic regime we can expand the l.h.s. before performing the integral, so we get
\be
\De E \ \geq \ \ga \frac{m^2}{|\G k|}\, ,
\ee
where $\ga \equiv -\frac{2 \pi \sqrt{2}}{4 \log \left(3-2 \sqrt{2}\right)} \approx 1.26$ .

On the other side, when $|{\bf{k}}| \ll m$, we can take $\om_\G k \approx m$ so that
\be
\De E \ \geq \  m \, .
\ee

The TEUR accurately captures a crucial fact often overlooked in textbooks, that usually focus on asymptotic theory: describing finite-time dynamics requires an intrinsic energy uncertainty of the particle states. 
 

\section{Conclusions}

In this paper we have developed a QFT framework for massive chiral fields in which particles (field modes) are described as carrying simultaneously well-defined chirality and helicity, rather than energy and helicity, as usual. This construction enables a consistent treatment of chiral dynamics beyond first-quantized approaches and provides a natural setting in which chiral oscillations emerge from the underlying field operators.
It also provides a consistent treatment of weak decay processes, which incorporate chiral oscillations and reduce to known results in the appropriate limits.

Our formulation takes advantage of previous studies on the quantization of fields with mass mixing \cite{Blasone:1995zc}, like neutrinos. Indeed, the chiral components $\psi_R$ and $\psi_L$ of a Dirac field can be seen as result of mixing between two fields with different masses, namely $\Psi$ and ${\widetilde \Psi}$, respectively with masses $m$ and $-m$. Thus, one finds the ladder operators for chiral particles, which turn out to be a generalization of recent results \cite{Bittencourt:2024yxi}. Also, the vacuum for the chiral fields turns out to be a condensate of particle-antiparticle pairs of the $\Psi$ and ${\widetilde \Psi}$ fields.

Building on this formulation, we derived chirality–energy uncertainty relations that quantify the interplay between the non-conservation of chirality for massive fermions and the spectral properties of the Hamiltonian. By identifying chiral charges as  ``clock observables'', we rephrased the chirality–energy uncertainty relations in terms of the Mandelstam--Tamm type of TEUR that are pertinent for the chiral sector. The latter offered a precise characterization of the time scales governing chiral transitions. The uncertainty relations thus obtained  provide a unified perspective in which chiral oscillations and uncertainty relations arise as complementary manifestations of the same underlying physics.

A distinctive feature of the present formulation with respect to the one presented in Ref.\cite{Bittencourt:2024yxi} is the appearance of the auxiliary field  ${\widetilde \Psi}$: this is indeed necessary for describing field modes with both combinations of helicities vs. chiralities. Although the ``tilde" field with negative mass does not enter in physical quantities, nevertheless its presence demands for a possible physical interpretation, which at present is still lacking. 

An important role in our discussion is played by contextuality \cite{Kochen:1967equ}. It is worth noting that, in high-energy physics, information about the measurement apparatus is associated to the interaction Lagrangian. For instance, the charged lepton produced in pion decay can be detected either through charged-current interactions or through neutral-current interactions, weak or electromagnetic. 
Using a detector based on  charged-current weak interaction, chiral oscillations can be observed. On the other hand, a detector based on electromagnetic interactions is not sensitive to chirality, and therefore   chiral oscillations cannot be observed in this way.
The situation is analogous to neutrino oscillations, which can be observed only if the detector operates through charged-current weak interactions. If neutrinos are detected via neutral-current interactions, which are insensitive to flavor, no neutrino disappearance is observed \cite{Chen:1985na,SNO:2002}.

A detailed analysis of the phenomenology of chiral oscillations based on the formalism here developed, will be presented elsewhere.







\section*{Acknowledgments}P.J. acknowledges support from
the Czech Science Foundation (GA\v{C}R), Grant No. 25-18105S. 

\appendix
\section{Pion decay revisited}

We consider the standard (helicity state) amplitude
\be
\ \lan 0| b^1_{\G p} \al^1_{\G k} \, J^{+\mu}_l(x)|0\ran \ = \ \ {}_{L,R}\lan 0(t_i)| b^1_{\G p} \al^1_{\G k,L}(t_i) \, \bar{G}^{-1}(t_i) J^{+\mu}_l(x) \bar{G}(t_i)|0(t_i)\ran_{L,R} \, . 
\ee
In order to proceed we need the Bogoliubov transformation
\begin{eqnarray}\label{chiraltransf1a2}
   \bar{G}^{-1}(t) \alpha_{{\bf k},R}^{1}(t) \bar{G}(t) &=&
    \frac{1}{\sqrt{2}} \alpha_{{\bf k},R}^{1}(t) \,+\,\frac{1}{\sqrt{2}}\left(\Uk\, \alpha_{{\bf k},L}^{1} (t)\,+\, \Vk\, \bt_{-{\bf k},L}^{1\dagger}(t) e^{2 i \om_k t}\right)
    \\ \label{chiraltransf1b2}
   \bar{G}^{-1}(t)  \bt_{-{\bf k},R}^{1\dagger}(t) \bar{G}(t) &=&
     \frac{1}{\sqrt{2}} \bt_{-{\bf k},R}^{1\dagger}(t) \,+\,\frac{1}{\sqrt{2}}\left(\Uk\,  \bt_{-{\bf k},L}^{1\dagger} (t) \,-\, \Vk\,
     \alpha_{{\bf k},L}^{1}(t) e^{-2 i \om_k t}
     \right)
     \\ \label{chiraltransf1c2}
      \bar{G}^{-1}(t)  \alpha_{{\bf k},L}^{1}(t) \bar{G}(t) &=&
    \frac{1}{\sqrt{2}} \alpha_{{\bf k},L}^{1}(t) \,-\,\frac{1}{\sqrt{2}}\left(\Uk\,\alpha_{{\bf k},R}^{1}(t) \,-\, \Vk\, \bt_{-{\bf k},R}^{1\dagger}(t) e^{2 i \om_k t}\right)
    \\ \label{chiraltransf1d2}
    \bar{G}^{-1}(t) \bt_{-{\bf k},L}^{1\dagger}(t) \bar{G}(t) &=&
     \frac{1}{\sqrt{2}}  \bt_{-{\bf k},L}^{1\dagger}(t) \,-\,\frac{1}{\sqrt{2}}\left(\Uk\, \bt_{-{\bf k},R}^{1\dagger} (t) \,+\, \Vk\,
    \alpha_{{\bf k},R}^{1}(t) e^{-2 i \om_k t}
     \right)
\end{eqnarray}
The above calculation can be performed at every $t_i$. We can choose $t_i=0$ for simplicity. Therefore, using the expansion \eqref{psiti}, we get
\be
\sqrt{2}{}_{L,R}\lan 0| b^1_{\G p} \al^1_{\G k,L} \, \bar{G}^{-1}(0) J^{+\mu}_l(x) \bar{G}(0)|0(0)\ran_{L,R} \ = \ e^{i (\G k+\G p) \cdot \G x} e^{i p t}\lf(U^{\G k,1 \dag}_{L L}(t,0)+\Uk \, U^{\G k,1 \dag}_{L R}(t,0)-\Vk \, V^{\G k,1 \dag}_{L R}(t,0) \ri) v_{\G p,L}
\ee
The factor $\sqrt{2}$ must be inserted because of Eq.\eqref{sq2} and the discussion below that equation.
One can verify that 
\be
 U^{\G k,1 }_{L L}(t,0)+\Uk \, U^{\G k,1}_{L R}(t,0)-\Vk \, V^{\G k,1}_{L R}(t,0) \ = \  u_\G k^1 e^{-i \om_k t} \, . 
\ee
Then
\be
\sqrt{2}{}_{L,R}\lan 0| b^1_{\G p} \al^1_{\G k,L} \, \bar{G}^{-1}(0) J^{+\mu}_l(x) \bar{G}(0)|0(0)\ran_{L,R} \ = \ e^{i (\G k+\G p) \cdot \G x} e^{i (p+\om_k) t} u^1_{\G k} v_{\G p,L} \, , 
\ee
which is the correct result.

\section{Quantum field theory of fermion mixing}

We report here, for convenience of the reader, a brief review of the quantization of mixed (neutrino) Dirac fields in the case of two flavors \cite{Blasone:1995zc}.

Mixing transformation 
\be
\begin{pmatrix} \nu_e(x) \\ \nu_\mu(x) \end{pmatrix} \ = \ \begin{pmatrix} \cos \theta & \sin \theta \\ -\sin \theta & \cos \theta \end{pmatrix}  \, \begin{pmatrix} \nu_1(x) \\ \nu_2(x) \end{pmatrix} \, ,
\ee
can be rewritten as
\bea  \non
\nu_{e}(x)  &=& G^{-1}_{\theta}(t)
\nu_{1}(x)
G_{\theta}(t) \\[2mm]
\nu_{\mu}(x) &=& G^{-1}_{\theta}(t)
\nu_{2}(x) \;
 G_{\theta}(t) \label{Gmix2}
\eea
with the generator given by:
\bea G_\theta(t) \ = \ \exp \lf[\theta \int d^{3}{\bf x}\, \lf(\nu_{1}^{\dag}(x) \, \nu_{2}(x)-\nu_{2}^{\dag}(x) \, 
\nu_{1}(x)\ri)\ri]\, .
\eea
This is obtained by means of the canonical anticommutation relations for the free Dirac fields $\nu_1$ and $\nu_2$.

The flavor vacuum is 
\be \label{timedep}
|0 (t)\ran_\flav \equiv G^{-1}_\theta(t)\; |0 \ran_\mass \, .
\ee
The state \eqref{timedep} is known as \emph{flavor vacuum} and it is annihilated by the operators $\al_{\sigma}(t)$ and $\bt_{\sigma}(t)$, given by\footnote{This form of the flavor operators is valid in a frame where $\G k=(0,0,|\G k|).$}:
\bea \lab{operat1}
\al^r_{{\bf k},e}(t)=\cos\theta\,\al^r_{{\bf k},1} +
\sin\theta \lf(
 U_{{\bf k}}^{*}(t)\, \al^r_{{\bf k},2}
 + \epsilon^r
V_{{\bf k}}(t)\, \bt^{r\dag}_{-{\bf k},2}\ri)\quad &&\\
\lab{operat2}
\al^r_{{\bf k},\mu}(t)=\cos\theta\,\al^r_{{\bf
k},2}-\sin\theta \lf(
 U_{{\bf k}}(t)\, \al^r_{{\bf k},1}
 - \epsilon^r
V_{{\bf k}}(t)\, \bt^{r\dag}_{-{\bf k},1}\ri)\quad  &&\\
\lab{operat3}
\!\!\bt^r_{-{\bf k},e}(t)=\cos\theta\,\bt^r_{-{\bf
k},1}+\sin\theta\lf(
U_{{\bf k}}^{*}(t)\, \bt^r_{-{\bf k},2}
 -\epsilon^r
V_{{\bf k}}(t)\, \al^{r\dag}_{{\bf k},2}\ri) \;\; &&\\ \lab{operat4}
\!\!\bt^r_{-{\bf k},\mu}(t)=\cos\theta \, \bt^r_{-{\bf k},2} -
\sin\theta \lf(
 U_{\bf k}(t)\, \bt^r_{-{\bf k},1}  + \epsilon^r
V_{{\bf k}}(t) \, \al^{r\dag}_{{\bf k},1} \ri) \;\; &&
\eea
In Eqs.(\ref{operat1})-(\ref{operat4}), $\epsilon^r \equiv (-1)^r$, and $U_{\bf k}\,$ and $\,V_{\bf k}$ are the \emph{mixing Bogoliubov
coefficients}:
\begin{eqnarray}
U_{{\bf k}}(t)& \equiv & u^{r\dag}_{{\bf k},2}u^r_{{\bf k},1}\;
e^{i(\om_{\G k,2}-\om_{\G k,1})t} \ = \ |U_\G k| \, e^{i(\om_{\G k,2}-\om_{\G k,1})t} \,  \, ,  \\[2mm]
V_{{\bf k}}(t) & \equiv & \epsilon^r\; u^{r\dag}_{{\bf k},1}v^r_{-{\bf k},2}\;
e^{i(\om_{\G k,2}+\om_{\G k,1})t} \ = \ |V_\G k| \, e^{i(\om_{\G k,2}+\om_{\G k,1})t} \, .
\end{eqnarray}
with
\bea \label{ucoff}
|U_\G k| & \equiv & u^{r\dag}_{{\bf k},2} \, u^{r}_{{\bf k},1} \ = \  v^{r\dag}_{-{\bf k},1} \, v^{r}_{-{\bf k},2} 
= \sqrt{\frac{(\omega_{\G k,1}+m_{1})(\omega_{\G k,2}+m_{2})}{4\omega_{\G k,1}\omega_{\G k,2}}} 
\left(1+\frac{{\bf k}^{2}}{(\omega_{\G k,1}+m_{1})(\omega_{\G k,2}+m_{2})}\right) \, . \\ [2mm] \label{vcoff}
|V_\G k| & = & \epsilon^r\; u^{r\dag}_{{\bf
k},1} \, v^{r}_{-{\bf k},2} \ = \  -\epsilon^r\, u^{r\dag}_{{\bf
k},2} \, v^{r}_{-{\bf k},1} 
= \frac{|\G k|}{\sqrt{4 \om_{\G k,1}\om_{\G k,2}}}
\lf(\sqrt{\frac{\om_{\G k,1}+m_1}{\om_{\G k,2}+m_2}}-\sqrt{\frac{\om_{\G k,2}+m_2}{\om_{\G k,1}+m_1}}\ri) \, .
\eea
and $
|U_{\bf k}|^2 + |V_{\bf k}|^2 \ = \ 1 \, .
$

The explicit for of the flavor vacuum is
\begin{eqnarray} \non
&& 
|0\ran_\flav= \prod_{{\bf k},r} \lf[
(1-\sin^2\theta\,|V_{{\bf
k}}|^2)
 \ri.-\,\epsilon^r\sin\theta\,\cos\theta\, |V_{{\bf k}}|
\, (\al^{r\dag}_{{\bf k},1}\bt^{r\dag}_{-{\bf k},2}+
\al^{r\dag}_{{\bf k},2} \bt^{r\dag}_{-{\bf k},1})\\\label{vacuumflav}
&&\lf.
+\,\epsilon^r\sin^2\theta \,|V_{{\bf k}}| |U_{{\bf k}}| \,(
\al^{r\dag}_{{\bf k},1}\bt^{r\dag}_{-{\bf k},1} -
\al^{r\dag}_{{\bf k},2}\bt^{r\dag}_{-{\bf k},2} )
+\,\sin^2\theta \, |V_{{\bf k}}|^2
\, \al^{r\dag}_{{\bf k},1}\bt^{r\dag}_{-{\bf k},2}
\al^{r\dag}_{{\bf k},2}\bt^{r\dag}_{-{\bf
k},1}\ri]|0\ran_\mass \, .
\end{eqnarray}

The chiral operators and Bogoliubov coefficients $\Uk$, $\Vk$ are obtained by setting $\theta \rightarrow \pi/4$, $m_1\rar - m$, $m_2\rightarrow m$, $e \rar R$, $\mu \rar L $. We then have $U_k\rar \Uk$ and $V_k\rar -\Vk $. This correspondence includes also the operators: $\alpha_1\rar {\widetilde \alpha}$, $\alpha_2\rar  \alpha$, etc.

\section{Useful formulas}

$$\gamma_{i}=\left(\begin{array}{cc} 0 & \sigma_{i} \\ - \sigma_{i} & 0
\end{array}\right)\qquad\quad
\gamma_{0}=\left(\begin{array}{cc} 1_{2} & 0 \\ 0 & -1_{2}
\end{array}\right)$$

\begin{equation}
    u_{k,i}^{r}=
A_{k,i}\;
\left(\begin{array}{c} \chi^r \\ \frac{k}{\omega_{i}+m_{i}} \chi^r\end{array}\right)\, , \qquad
v_{k,i}^{r}=A_{k,i}\;
\left(\begin{array}{c} \frac{k}{\omega_{i}+m_{i}}\chi^r \\  \chi^r\end{array}\right)
\end{equation}

where $A_{k,i}\equiv \left(\frac{\omega_{i}+m_{i}}{2\omega_{i}}\right)^{\frac{1}{2}}$ and $\chi^1=\left(\begin{array}{c} 1 \\ 0\end{array}\right)$, $\chi^2=\left(\begin{array}{c} 0 \\ 1\end{array}\right)$.

Calculations are most easily done in the reference frame wher $\G k=(0,0,|\G k|)$. The explicit form of the spinors reads:
$$u_{\G k,i}^{1}=
\left(\frac{\omega_{i}+m_{i}}{2\omega_{i}}\right)^{\frac{1}{2}}\;
\left(\begin{array}{c} 1 \\ 0 \\ \frac{|\G k|}{\omega_{i}+m_{i}} \\
 0\end{array}\right)
\; \;\;\;\;\;\;\;\;\;\;\;\;\;\;\;\;\;\; \;\;\;\;\;\;
u_{\G k,i}^{2}=
\left(\frac{\omega_{i}+m_{i}}{2\omega_{k,i}}\right)^{\frac{1}{2}}\;
\left(\begin{array}{c} 0 \\ 1 \\ 0  \\
 \frac{-|\G k|}{\omega_{k,i}+m_{i}}\end{array}\right)$$

$$v_{-\G k,i}^{1}=
\left(\frac{\omega_{i}+m_{i}}{2\omega_{k,i}}\right)^{\frac{1}{2}}\;
\left(\begin{array}{c}  \frac{-|\G k|}{\omega_{k,i}+m_{i}} \\
 0 \\ 1 \\ 0\end{array}\right)
\; \;\;\;\;\;\;\;\;\;\;\;\;\;\;\;\;\;\; \;\;\;\;\;\;
v_{-\G k,i}^{2}=
\left(\frac{\omega_{i}+m_{i}}{2\omega_{k,i}}\right)^{\frac{1}{2}}\;
\left(\begin{array}{c}  0 \\
 \frac{|\G k|}{\omega_{k,i}+m_{i}} \\ 0 \\ 1\end{array}\right)$$

\section{Consistency check of left and right field expansions}

By means of  relations \eqref{uvtilde1}, \eqref{uvtilde2} we obtain the following identities:
\begin{eqnarray}
    {\widetilde u}_{{\bf k}}^{1} &=&  u_{{\bf k}}^{1}U_k+ v_{{-\bf k}}^{1}V_k
    \\
    {\widetilde v}_{-{\bf k}}^{1} &=&  -u_{{\bf k}}^{1}V_k+ v_{{-\bf k}}^{1}U_k
    \\
    {\widetilde u}_{{\bf k}}^{2} &=&  u_{{\bf k}}^{2}U_k- v_{{-\bf k}}^{2}V_k
    \\
    {\widetilde v}_{-{\bf k}}^{2} &=&  u_{{\bf k}}^{2}V_k+ v_{{-\bf k}}^{2}U_k
\end{eqnarray}
and
\begin{eqnarray}
    u_{{\bf k}}^{1} &=&  {\widetilde u}_{{\bf k}}^{1} U_k- {\widetilde v}_{{-\bf k}}^{1}V_k
    \\
    u_{{\bf k}}^{2} &=&  + {\widetilde u}_{{\bf k}}^{2}U_k + {\widetilde v}_{{-\bf k}}^{2}V_k
    \\
    v_{-{\bf k}}^{2} &=&  
 - {\widetilde u}_{{\bf k}}^{2}V_k + {\widetilde v}_{{-\bf k}}^{2}U_k
    \\
    v_{-{\bf k}}^{1} &=& {\widetilde u}_{{\bf k}}^{1}V_k + {\widetilde v}_{{-\bf k}}^{1}U_k
\end{eqnarray}

These can be used to check the consistency of the field expansions of $\psi_{L,R}$ in terms of the chiral ladder operators. 

Let us consider $\psi_R$. We have (at time $t=0$)
\begin{eqnarray} \nonumber
\psi_R &=&  \int \Big[ 
 {\widetilde u}_{{\bf k}}^{1} \,\alpha_{{\bf k},R}^{1} 
 + 
\, {\widetilde u}_{{\bf k}}^{2} \,\alpha_{{\bf k},R}^{2} 
+ 
 {\widetilde v}_{-{\bf k}}^{2} \,\beta_{-{\bf k},R}^{2\dagger} 
 + 
 {\widetilde v}_{-{\bf k}}^{1} \beta_{-{\bf k},R}^{1\dagger} 
\Big]  
\end{eqnarray}
By substituting the chiral operators and collecting with respect to the helicity operators, we obtain:
\begin{eqnarray}  
\psi_R &=&   \frac{1}{\sqrt{2}}\int \Big[ 
 {\widetilde u}_{{\bf k}}^{1}
 {\widetilde \alpha}_{{\bf k}}^{1}
 +
{\widetilde u}_{{\bf k}}^{2}
 {\widetilde \alpha}_{{\bf k}}^{2}
+ 
 \left({\widetilde u}_{{\bf k}}^{1} U_k - {\widetilde v}_{-{\bf k}}^{1} V_k
 \right)\,\alpha_{{\bf k}}^{1} 
 + 
 \left({\widetilde u}_{{\bf k}}^{2} U_k +
 {\widetilde v}_{-{\bf k}}^{2} V_k
 \right)\,\alpha_{{\bf k}}^{2} 
 \\ 
&&+ {\widetilde v}_{-{\bf k}}^{1}
{\widetilde \beta}_{{\bf k}}^{1\dag} 
+{\widetilde v}_{-{\bf k}}^{2}
{\widetilde \beta}_{{\bf k}}^{2\dag} 
+\left( {\widetilde u}_{{\bf k}}^{1} V_k + {\widetilde v}_{-{\bf k}}^{1} U_k \right)
\beta_{{\bf k}}^{1\dag} 
+\left(- {\widetilde u}_{{\bf k}}^{2} V_k + {\widetilde v}_{-{\bf k}}^{2} U_k \right)
\beta_{{\bf k}}^{2\dag}
\Big].
\end{eqnarray}
Using the above relations, we see that indeed $\Psi_R = ({\widetilde \Psi} + \Psi)/\sqrt{2}$.
A similar check can be done for $\psi_L$.

%
%

\bibliography{LibraryNeutrino}

\bibliographystyle{apsrev4-2}

\end{document}